\documentclass[aps,showkeys,twocolumn,showpacs,preprintnumbers,amsmath,amssymb,superscriptaddress,floatfix,nofootinbib]{revtex4}
\usepackage{graphicx,color,dcolumn,booktabs,bm}
\usepackage{longtable,lscape}
\usepackage{txfonts}
\usepackage{overpic}
\usepackage{epsfig}
\usepackage{amssymb}
\usepackage{rotating}
\usepackage{epstopdf}
\usepackage{indentfirst}
\usepackage{feynmf}   
\usepackage{slashed}  
\usepackage{cases}
\usepackage{color}
\usepackage{multirow}
\usepackage{graphicx,color,dcolumn,booktabs,bm}
\usepackage{cases}
\usepackage{array}

\begin{document}

\title{The $X(4140)$, $X(4274)$, $X(4500)$ and $X(4700)$ in the relativized quark model}

\author{Qi-Fang L\"{u}} \email{lvqifang@ihep.ac.cn}
\affiliation{Institute of High Energy Physics, Chinese Academy of Sciences, Beijing 100049, China}
\author{Yu-Bing Dong} \email{dongyb@ihep.ac.cn}
\affiliation{Institute of High Energy Physics, Chinese Academy of Sciences, Beijing 100049, China}
\affiliation{Theoretical Physics Center for Science Facilities (TPCSF), CAS, Beijing 100049, China}

\begin{abstract}

We investigate the masses of $cs \bar c \bar s$ tetraquark states in a diquark-antidiquark picture employing the relativized quark model proposed by Godfrey and Isgur. Only the antitriplet diquark states in color space are calculated. The diquark masses are obtained with the relativized potential firstly, and then the diquark and antidiquark are treated as the usual antiquark and quark, respectively, and the masses of the tetraquark states are obtained by solving the Schr\"{o}dinger equation with the relativized potential between the diquark and antidiquark. The theoretical uncertainties induced by screening effects are also taken into account. It is found that the resonance of $X(4140)$ can be regarded as the $cs \bar c \bar s$ tetraquark ground states, and the $X(4700)$ can be assigned as the $2S$ excited tetraquark state. When the internal excited diquarks are taken into account, the resonance of $X(4500)$ can be explained as the tetraquark composed of one $2S$ scalar diquark and one scalar antidiquark. In our approach, the $X(4274)$ cannot be explained as a tetraquark state, however, it can be a good candidate of the conventional $\chi_{c1}(3^3P_1)$ state. In addition, other charmonium-like states $\chi_{c0}(3915)$, $X(4350)$, $X(4630)$ and $X(4660)$, as the $cs \bar c \bar s$ tetraquark states, are also discussed.

\end{abstract}
\maketitle

\section{Introduction}{\label{introduction}}

Very recently, the LHCb Collaboration has performed the first full amplitude analysis of $B^+ \to J/\psi \phi K^+$ process with $pp$ collision data collected at $\sqrt{s} = 7$ and 8 TeV~\cite{Aaij:2016iza,Aaij:2016nsc}. Besides the confirmation of the two resonances $X(4140)$ and $X(4274)$ in the $J/\psi \phi$ invariant mass, two new structures $X(4500)$ and $X(4700)$ are also observed at the same time. The spin parities are $1^{++}$ for the $X(4140)$ and $X(4274)$, and $0^{++}$ for the $X(4500)$ and $X(4700)$. The measured masses and total decay widths are
\begin{eqnarray}
(M; \Gamma)_{X(4140)} = (4146.5\pm4.5^{+4.6}_{-2.8};~ 83\pm21^{+21}_{-14})~\rm{MeV},
\end{eqnarray}
\begin{eqnarray}
(M; \Gamma)_{X(4274)} = (4273.3\pm8.3^{+17.2}_{-3.6};~ 56\pm11^{+8}_{-11})~\rm{MeV},
\end{eqnarray}
\begin{eqnarray}
(M; \Gamma)_{X(4500)} = (4506\pm11^{+12}_{-15};~ 92\pm21^{+21}_{-20})~\rm{MeV},
\end{eqnarray}
\begin{eqnarray}
(M;\Gamma)_{X(4700)} = (4704\pm10^{+14}_{-24};~120\pm31^{+42}_{-33})~\rm{MeV}.
\end{eqnarray}

It should be mentioned that the $X(4140)$ was first reported by the CDF Collaboration in the $J/\psi \phi$ invariant mass distribution of the $B^+ \to J/\psi \phi K^+$ decay in 2009~\cite{Aaltonen:2009tz}, and then this structure was observed by several collaborations in the next few years~\cite{Brodzicka:2010zz,Aaltonen:2011at,Aaij:2012pz,Chatrchyan:2013dma,Abazov:2013xda,Lees:2014lra,Abazov:2015sxa}. In 2011, the CDF Collaboration found the evidence of the $X(4274)$ with approximate significance of $3.1\sigma$~\cite{Aaltonen:2011at}. The related peaks of $J/\psi \phi$ mass structures around 4.3 GeV were also reported by LHCb, CMS, D0 and BaBar Collaborations~\cite{Aaij:2012pz,Chatrchyan:2013dma,Abazov:2013xda,Lees:2014lra}, which may be the same state as the $X(4274)$. It should be noted that the Belle Collaboration measured a narrow $J/\psi \phi$ peak of $X(4350)$ in the double photon collisions~\cite{Shen:2009vs}, which indicates the $X(4350)$ should have the spin parity $J^{PC}=0^{++}$ or $2^{++}$ and be a different structure from the $X(4274)$.

Various theoretical studies on the resonances of $X(4140)$ and $X(4274)$, such as the molecular states~\cite{Liu:2009ei,Mahajan:2009pj,Wang:2009ry,Wang:2009ue,Albuquerque:2009ak,Branz:2009yt,Ding:2009vd,Zhang:2009st,Shen:2010ky,Liu:2010hf,Finazzo:2011he,He:2011ed,HidalgoDuque:2012pq,Wang:2014gwa,Ma:2014ofa,Ma:2014zva,Karliner:2016ith,Torres:2016oyz}, compact or diquark-antidiquark states~\cite{Stancu:2009ka,Drenska:2009cd,Ebert:2010zz,Chen:2010ze,Vijande:2014cfa,Patel:2014vua,Anisovich:2015caa,Wang:2015pea,Zhou:2015frp,Padmanath:2015era,Lebed:2016yvr}, cusp effects~\cite{vanBeveren:2009dc,Swanson:2014tra}, dynamically generated resonances~\cite{Molina:2009ct,Branz:2010rj}, conventional charmonium~\cite{Liu:2009iw}, and hybrid charmonium states~\cite{Mahajan:2009pj,Wang:2009ue}, have been performed in the literature. Given the $J^{PC}=1^{++}$ for the both states, many molecular and hybrid charmonium interpretations with other quantum numbers can be ruled out~\cite{Liu:2009ei,Albuquerque:2009ak,Branz:2009yt,Ding:2009vd,Zhang:2009st,Mahajan:2009pj,Wang:2009ue}. It should be noticed that the cusp effects may explain the structure of the $X(4140)$, but fail to account for the $X(4274)$~\cite{Swanson:2014tra}. Moreover, the compact tetraquark model, implemented by Stancu, can describe the $X(4140)$ and $X(4274)$ simultaneously~\cite{Stancu:2009ka}, while only one $J^{PC}=1^{++}$ state exists in the color triplet diquark-antidiquark picture in this energy region~\cite{Ebert:2010zz,Anisovich:2015caa,Lebed:2016yvr}. A comprehensive overview about the $X(4140)$ and $X(4274)$ can be found in Ref.~\cite{Chen:2016qju}.

After the new observations of the LHCb Collaboration, several theoretical works have been proposed. By using the QCD sum rule, the $X(4140)$ and $X(4274)$ were interpreted as the $S$-wave $c\bar c s\bar s$ tetraquark states with opposite color structures, and the $X(4500)$ and $X(4700)$ as the $D$-wave $c\bar c s\bar s$ tetraquark states also with opposite color structures~\cite{Chen:2016oma}. In Refs.~\cite{Wang:2016gxp,Wang:2016tzr}, the $X(4500)$ is assigned as the first radial excited axial-vector diqaurk and axial-vector-antidiquark type tetraquark, the $X(4700)$ was assigned as the tetraquark ground state with a vector-diquark and a vector-antidiquark, and the $X(4140)$ is disfavored as a $c\bar c s\bar s$ tetraquark state. The possible re-scattering effects in the $B^+\to J/\psi \phi K^+$ are also investigated, which shows that those effects may simulate the structures of the $X(4140)$ and $X(4700)$, but hardly explain the $X(4272)$ and $X(4500)$~\cite{Liu:2016onn}. Based on the spin-spin interaction, Maiani et al., suggest that the $X(4500)$ and $X(4700)$ are the $2S$ $cs \bar c \bar s$ tetraquark state, the $X(4140)$ is the ground state, and the $X(4274)$ may have quantum number $0^{++}$ or $2^{++}$~\cite{Maiani:2016wlq}. Moreover, a detailed calculation is performed by Zhu~\cite{Zhu:2016arf} where the $X(4140)$ and $X(4274)$ may be described simultaneously by adding the up and down quark components. Other studies related with these resonances are also discussed~\cite{Ali:2016dkf,He:2016pfa,Wang:2016ujn}. It should be stressed that all those interpretations do not agree with each other, and most of them are obtained with the QCD sum rule and the quark model with only spin-spin interactions. Hence, it is essential to study the four resonance structures, especially the $X(4500)$ and $X(4700)$, in a realistic potential model for a comparison.

In this work, we employ the relativized quark model proposed by Godfrey and Isgur (GI model) to calculate the masses of $cs$ diquark and $cs\bar c \bar s$ tetraquark states. It should be mentioned that this model has been widely used to calculate the masses of the conventional mesons and baryons~\cite{Godfrey:1985xj,Capstick:1986bm,Godfrey:1998pd,Ferretti:2013faa,Ferretti:2013vua,Godfrey:2015dva}. The obtained wave functions are also employed to estimate their strong decay widths, radiations, decay constants, and leptonic decays ~\cite{Capstick:1992th,Barnes:2005pb,Liu:2009fe,Zhou:2013ada,Lu:2014zua,Song:2015nia,Song:2015fha,Godfrey:2014fga,Godfrey:2015dia,Ferretti:2015rsa,Capstick:1989ra,Yaouanc:2014jka,Godfrey:2016nwn}. It is believed that this model can give a unified description of the low lying mesons and baryons, and therefore, it is suitable to deal with the $cs\bar c \bar s$ tetraquark states, where
both heavy-light and heavy-heavy systems are included. To calculate the tetraquark masses, we restrict present calculations in the diquark-antidiquark picture with color triplet following the route proposed by Ebert, Faustov, and Galkin~\cite{Ebert:2002ig,Ebert:2005nc,Ebert:2008kb,Ebert:2010af,Santopinto:2014opa,Hadizadeh:2015cvx}. First of all, the corresponding diquark and antidiquark masses are estimated with this relativized potential. Then, they are treated as the usual point-like antiquark and quark, respectively. The masses of tetraquark states are, therefore,  obtained by solving the Schr\"{o}dinger-type equation between diquark and antidiquark. This method has been used to discuss the $X(5568)$ observed by D0 Collaboration in our previous work~\cite{D0:2016mwd,LHCb:2016ppf,Lu:2016zhe}, and should be appropriate to investigate the abundant charmonium-like states. Moreover, the color screening effects of the confinement are also considered here. We find that the $X(4140)$ can be assigned as the ground tetraquark state, and the $X(4500)$ and $X(4700)$ are the good candidates of the excited tetraquarks. There is no room left for the $X(4274)$ in the color triplet diquark-antidiquark picture, however, the mass and total decay width of the $X(4274)$ are consistent with the conventional $\chi_{c1}(3^3P_1)$ in the relativized quark model.

This paper is organized as follows.  The relativized quark model and the screening potential of confinement are briefly introduced, and the masses of $cs$ diquarks are calculated in Sec.~\ref{diquark}. In Sec.~\ref{tetraquark}, the masses of $cs\bar c \bar s$ tetraquark states and $P$ wave charmonium are numerically estimated, and discussions are also presented. Finally, we give a short summary of our study in the last section.

\section{Relativized Quark Model and Masses of $cs$ Diquarks}{\label{diquark}}

The Hamiltonian between the quark and antiquark in the relativized quark model can be expressed as
\begin{equation}
\tilde{H} = H_0+\tilde{V}(\boldsymbol{p},\boldsymbol{r}), \label{ham}
\end{equation}
with
\begin{equation}
H_0 = (p^2+m_1^2)^{1/2}+(p^2+m_2^2)^{1/2},
\end{equation}
\begin{equation}
\tilde{V}(\boldsymbol{p},\boldsymbol{r}) = \tilde{H}^{conf}_{12}+\tilde{H}^{cont}_{12}+\tilde{H}^{ten}_{12}+\tilde{H}^{so}_{12},
\end{equation}
where the $\tilde{H}^{conf}_{12}$ includes the spin-independent linear confinement and Coulomb-like interaction, the $\tilde{H}^{cont}_{12}$, $\tilde{H}^{ten}_{12}$, and $\tilde{H}^{so}_{12}$ are the color contact term, the color tensor interaction, and the spin-orbit term, respectively. The $\tilde{H}$ represents that the operator $H$ has taken account of the relativistic effects via the relativized procedure. The explicit forms of those interactions and the details of the relativization procedure can be found in Appendix A of Ref.~\cite{Godfrey:1985xj}.

Since the GI model is a typical quenched quark model, the coupled channel effects or the screening effects have been ignored. Those effects may influence on the excited mass spectrum of mesons, for example the leptonic decay rates of charmonium~\cite{Dong:1994zj,Ding:1995he} and the lower mass puzzle of $D^*_{s0}(2317)$ and $D_{s1}(2460)$~\cite{Song:2015nia}. The unquenched properties are reflected by replacing the linear confinement to the screening potential, that is, the $br \to b(1-e^{-\mu r})/\mu$, where the $b$ denotes the string tension and the $\mu$ is the screen parameter~\cite{Song:2015nia,Song:2015fha}. The modified formalism with a new screening parameter gives a better description of the charmed and charmed-strange meson spectra. Besides the ground $cs\bar c \bar s$ tetraquark states, the excited ones are also needed to describe the newly observed resonances. Here, we stress that when one deals with the $cs$ diquarks and $cs\bar c \bar s$ tetraquarks, the screening effects should also be considered for completeness.

In present work, only the antitriplet diquark $[\bar 3_c]_{cs}$ are considered. It should be noticed that the $[6_c]_{cs}$ type diquarks can not be formed in the GI quark model, since the confinement becomes repulsive in this case. For the quark-quark interaction in the antitriplet diquark system, the relation of $\tilde{V}_{cs}(\boldsymbol{p},\boldsymbol{r})=\tilde{V}_{c\bar s}(\boldsymbol{p},\boldsymbol{r})/2$ is employed. The model parameters employed in our calculations are the same as the ones in the original work~\cite{Godfrey:1985xj}. The screening parameter $\mu$ varies from 0 to 0.04 GeV, where the $\mu \to 0$ case is equivalent to the linear confinement $br$, and the $\mu = 0.02~\rm{GeV}$ case can improve the description of the charmed-strange spectrum significantly~\cite{Song:2015nia,Song:2015fha}.

Conventionally, the diquarks are classified into two groups: the ground states locating in $1S$ wave and the ones with internal excitations. For the diquarks lying in the ground states, the spin-parities are $J^P=0^+$ and $J^P=1^+$, named as scalar diquark and axial-vector one, respectively. For the diquarks with internal excitations, we only consider the $1P$, $2S$ and $1D$ waves in this work, and restrict the total angular momentum $J=0$ or 1. With the spectrum notation, the excited diquarks can be respectively denoted as $1^1P_1$, $1^3P_0$, $1^3P_1$, $2^1S_0$, $2^3S_1$, and $1^3D_1$. Here, we use the Gaussian expansion method to solve the Hamiltonian~(\ref{ham}) with $\tilde{V}_{cs}(\boldsymbol{p},\boldsymbol{r})$ potential~\cite{Hiyama:2003cu}. The obtained masses of the $cs$ diquarks are presented in Table.~\ref{tab1}. It can be seen that the diquark masses decreases with the screening effects considered. While the $\mu$ varies from 0 to 0.02 GeV, the diquark masses change about 10 MeV for ground states, and $20\sim35~\rm{MeV}$ for the excited states. Since the $\mu=0.02~\rm{GeV}$ case can give a better description of the $c\bar s$ meson spectra~\cite{Song:2015nia}, we prefer to adopt the diquark masses at this value to calculate the tetraquark states, and present the $\mu \to 0$ and $0.04~\rm{GeV}$ cases as the theoretical uncertainties.

\begin{table}[!htbp]
\begin{center}
\caption{ \label{tab1} Obtained masses of the $cs$ diquarks. $S$ and $A$ denote scalar and axial-vector diquarks in the ground states, respectively. The notation $n^{2S+1}P_J$ is used to stand for the excited diquarks. The brace and bracket correspond to symmetric and antisymmetric quark contents in flavor, respectively. The unit is in MeV.}
\small
\begin{tabular*}{8.5cm}{@{\extracolsep{\fill}}*{5}{p{1.5cm}<{\centering}}}
\hline\hline
 Quark &  Diquark   & Mass   & Mass $(\mu=$ & Mass $(\mu=$ \\
 content &   type   & (GI model)   & $0.02~\rm{GeV}$)  & $0.04~\rm{GeV}$)\\\hline
 $[c,s]$       &  $S$            & 2230       & 2221              & 2212                 \\
 $\{c,s\}$     &  $A$            & 2264       & 2254              & 2244                 \\
 $[c,s]$       &  $1^1P_1$       & 2523       & 2503              & 2482                 \\
 $\{c,s\}$     &  $1^3P_0$       & 2518       & 2496              & 2475                 \\
 $[c,s]$       &  $1^3P_1$       & 2529       & 2508              & 2486                 \\
 $[c,s]$       &  $2^1S_0$       & 2624       & 2593              & 2563                 \\
 $\{c,s\}$     &  $2^3S_1$       & 2644       & 2612              & 2580                 \\
 $\{c,s\}$     &  $1^3D_1$       & 2743       & 2708              & 2673                 \\

 \hline\hline
\end{tabular*}
\end{center}
\end{table}

\section{Masses of $cs \bar c \bar s$ tetraquark states and $P$ wave charmonium}{\label{tetraquark}}

\subsection{Tetraquarks composed of $S$ and $A$ diquarks}

In this work, a diquark is treated as a point-like antitriple state or we assume the distance between diquark and antidiquark is large enough~\cite{Maiani:2004vq,Santopinto:2004hw,Ferretti:2011zz,Santopinto:2014opa,Maiani:2014aja,Maiani:2015vwa,Brodsky:2014xia,Lebed:2015tna}. Therefore, we restrict our calculations of low lying tetraquark states within the $N \equiv 2(n+n_d+n_{\bar d})+L+L_d+L_{\bar d} \leq 2$ shell, where the $n(L)$, $n_d(L_d)$, and $n_{\bar d}(L_{\bar d})$ are the radial (orbital) quantum numbers of the relative motion, the diquark, and the antidiquark, respectively.

\begin{figure}[!htbp]
\includegraphics[scale=0.52]{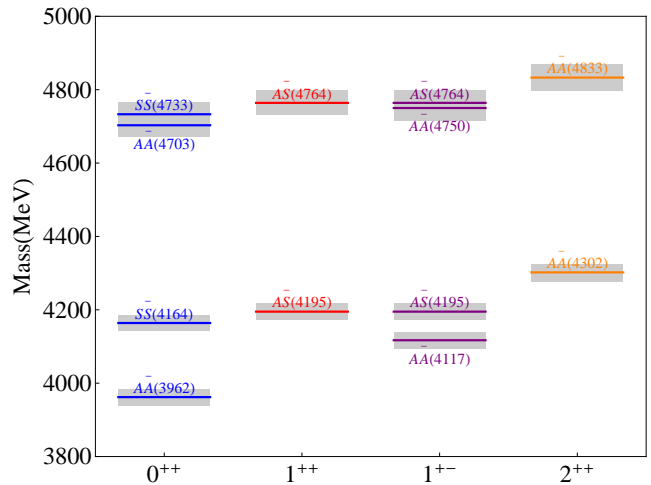}
\vspace{0.0cm} \caption{The predicted mass spectra of the $cs\bar c \bar s$ tetraquarks in $1S$ and $2S$ wave.}
\label{tetra}
\end{figure}

First of all, we calculate the tetraquark ground states composed of the $S$ and $A$ type diquarks, where the $l_d=l_{\bar d} =0$. Their mass spectra are listed in Tab.~\ref{mass1}. The $1S$ and $2S$ wave tetraquark states are also presented in Fig.~\ref{tetra}, where the central values are taken at $\mu=0.02~\rm{GeV}$ and the theoretical uncertainties are shown by the shaded bands. The theoretical errors induced by the screen effects are about $20\sim30$ MeV. It can be seen that the lowest state is the $J^P=0^+$ $A\bar A$ type diquark-antidiquark configuration rather than the $S\bar S$ type. This order of mass spectrum is different from the results of Refs.~\cite{Ebert:2005nc,Ebert:2008kb}, in which the $J^P=0^+$ $S\bar S$ type is the lowest one. We know that there exist fine splitting for the $A\bar A$ states via the spin-spin interaction.  The coefficients is -2, -1, and 2 for the $0^+$, $1^+$, and $2^+$ $A\bar A$ states respectively, if the spin-spin interaction is treated perturbatively. Although the mass of $A$ type diquark is higher than that of $S$ type, the larger fine splitting caused by the spin-spin interaction can suppress $J^P=0^+$ $A\bar A$ state to be the lowest one. The lowest mass of the $0^{++}$ state is 3962 MeV, which is consistent with the $\chi_{c0}(3915)$. Hence, it is possible to assign the $\chi_{c0}(3915)$ as the lightest $cs\bar c \bar s$ state, which has been proposed in Ref.~\cite{Lebed:2016yvr}.

\begin{table*}[!htbp]
\begin{center}
\caption{ \label{mass1} Masses of $cs\bar c \bar s$ tetraquark states composed of the $S$ and $A$ diquarks and antidiquarks in $1S$, $1P$, $2S$, and $1D$ waves. In the $A\bar S$ case, the linear combinations together with $S\bar A$ are understood to form the eigenstates of charge conjugation~\cite{Lebed:2016yvr}. The unit is in MeV.}
\small
\begin{tabular*}{17cm}{@{\extracolsep{\fill}}*{9}{p{1.7cm}<{\centering}}}
\hline\hline
 $J^{PC}$                &  Diquark        & Anti-diquark          & $n+1$           & $S$     & $L$     & Mass (GI model)  & Mass ($\mu=0.02~\rm{GeV}$) &  Mass ($\mu=0.04~\rm{GeV}$) \\\hline
 $|0^{++}\rangle$      &  $S$              & $\bar S$             & 1           & 0     & 0     & 4185             &  4164                 & 4143    \\
 $|0^{++}\rangle$      &  $A$              & $\bar A$             & 1           & 0     & 0     & 3984             &  3962                 & 3940    \\
 $|1^{++}\rangle$        &  $A$              & $\bar S$             & 1           & 1     & 0     & 4217           &  4195                 & 4173    \\
 $|1^{+-}\rangle$      &  $A$              & $\bar S$             & 1           & 1     & 0     & 4217             &  4195                 & 4173    \\
 $|1^{+-}\rangle$      &  $A$              & $\bar A$             & 1           & 1     & 0     & 4139             &  4117                 & 4095    \\
 $|2^{++}\rangle$        &  $A$              & $\bar A$             & 1           & 2     & 0     & 4325           &  4302                 & 4278    \\\hline
 $|0^{--}\rangle$        &  $A$              & $\bar S$             & 1           & 1     & 1     & 4599             &  4572                 & 4545    \\
 $|0^{-+}\rangle$      &  $A$              & $\bar S$             & 1           & 1     & 1     & 4599             &  4572                 & 4545    \\
 $|0^{-+}\rangle$      &  $A$              & $\bar A$             & 1           & 1     & 1     & 4595             &  4567                 & 4540    \\
 $|1^{--}\rangle$      &  $A$              & $\bar S$             & 1           & 1     & 1     & 4633             &  4605                 & 4578    \\
 $|1^{--}\rangle$      &  $S$              & $\bar S$             & 1           & 0     & 1     & 4632             &  4604                 & 4577    \\
 $|1^{-+}\rangle$      &  $A$              & $\bar S$             & 1           & 1     & 1     & 4633             &  4605                 & 4578    \\
 $|1^{-+}\rangle$      &  $A$              & $\bar A$             & 1           & 1     & 1     & 4680             &  4651                 & 4622    \\
 $|1^{--}\rangle$      &  $A$              & $\bar A$             & 1           & 0     & 1     & 4679             &  4651                 & 4622    \\
 $|1^{--}\rangle$      &  $A$              & $\bar A$             & 1           & 2     & 1     & 4599             &  4571                 & 4543    \\
 $|2^{-+}\rangle$      &  $A$              & $\bar S$             & 1           & 1     & 1     & 4691             &  4662                 & 4633    \\
 $|2^{-+}\rangle$      &  $A$              & $\bar A$             & 1           & 1     & 1     & 4706             &  4677                 & 4648    \\
 $|2^{--}\rangle$      &  $A$              & $\bar S$             & 1           & 1     & 1     & 4691             &  4662                 & 4633    \\
 $|2^{--}\rangle$      &  $A$              & $\bar A$             & 1           & 2     & 1     & 4702             &  4673                 & 4643    \\
 $|3^{--}\rangle$      &  $A$              & $\bar A$             & 1           & 2     & 1     & 4735             &  4705                 & 4675    \\\hline
 $|0^{++}\rangle$      &  $S$              & $\bar S$             & 2           & 0     & 0     & 4767             & 4733                  & 4700    \\
 $|0^{++}\rangle$      &  $A$              & $\bar A$             & 2           & 0     & 0     & 4736             & 4703                  & 4671    \\
 $|1^{++}\rangle$      &  $A$              & $\bar S$             & 2           & 1     & 0     & 4798             & 4764                  & 4730    \\
 $|1^{+-}\rangle$      &  $A$              & $\bar S$             & 2           & 1     & 0     & 4798             & 4764                  & 4730    \\
 $|1^{+-}\rangle$      &  $A$              & $\bar A$             & 2           & 1     & 0     & 4784             & 4750                  & 4716    \\
 $|2^{++}\rangle$      &  $A$              & $\bar A$             & 2           & 2     & 0     & 4870             & 4833                  & 4797    \\\hline
 $|0^{++}\rangle$      &  $A$              & $\bar A$             & 1           & 2     & 2     & 4949             & 4912                  & 4876    \\
 $|1^{++}\rangle$      &  $A$              & $\bar S$             & 1           & 1     & 2     & 4931             & 4896                  & 4860    \\
 $|1^{++}\rangle$      &  $A$              & $\bar A$             & 1           & 2     & 2     & 4962             & 4925                  & 4881    \\
 $|1^{+-}\rangle$      &  $A$              & $\bar S$             & 1           & 1     & 2     & 4931             & 4896                  & 4860    \\
 $|1^{+-}\rangle$      &  $A$              & $\bar A$             & 1           & 1     & 2     & 4969             & 4932                  & 4896    \\
 $|2^{++}\rangle$      &  $S$              & $\bar S$             & 1           & 2     & 2     & 4930             & 4895                  & 4859    \\
 $|2^{++}\rangle$      &  $A$              & $\bar S$             & 1           & 1     & 2     & 4952             & 4916                  & 4879    \\
 $|2^{++}\rangle$      &  $A$              & $\bar A$             & 1           & 0     & 2     & 4989             & 4952                  & 4915    \\
 $|2^{++}\rangle$      &  $A$              & $\bar A$             & 1           & 2     & 2     & 4982             & 4945                  & 4907    \\
 $|2^{+-}\rangle$      &  $A$              & $\bar S$             & 1           & 1     & 2     & 4952             & 4916                  & 4879    \\
 $|2^{+-}\rangle$      &  $A$              & $\bar A$             & 1           & 1     & 2     & 4992             & 4954                 & 4917    \\
 $|3^{++}\rangle$      &  $A$              & $\bar S$             & 1           & 1     & 2     & 4980             & 4943                  & 4906    \\
 $|3^{++}\rangle$      &  $A$              & $\bar A$             & 1           & 2     & 2     & 5001             & 4963                  & 4926    \\
 $|3^{+-}\rangle$      &  $A$              & $\bar S$             & 1           & 1     & 2     & 4980             & 4943                  & 4906    \\
 $|3^{+-}\rangle$      &  $A$              & $\bar A$             & 1           & 1     & 2     & 4999             & 4961                  & 4924    \\
 $|4^{++}\rangle$      &  $A$              & $\bar A$             & 1           & 2     & 2     & 5009             & 4972                  & 4934    \\
\hline\hline
\end{tabular*}
\end{center}
\end{table*}

The other obvious feature of our results is that the mass gap between the $1S$ $0^{++}$ doublet is larger, while the gap between the $2S$ states is extremely small and the theoretical errors overlap with each other. In the charmed mesons and charmonium sector, the splitting between the $2S$ states is much smaller than the $1S$ doublet~\cite{Agashe:2014kda} as well as the predicted charmed-strange meson spectrum~\cite{Godfrey:1985xj,Song:2015nia}. Since the spin-spin splitting is smaller for the $2S$ states, the lowest $2S$ $A\bar A$ with coefficient -2 becomes higher and close to the $2S$ $S\bar S$ state.  Their predicted masses are 4703 and 4733 MeV, which are in good agreement with the $X(4700)$. The obtained two $2S$ $0^{++}$ states are too close to distinguish by the masses and more information about $X(4700)$ are needed. In Ref.~\cite{Maiani:2016wlq}, the $X(4500)$ and $X(4700)$ are attributed to the $2S$ doublet. However, the 200 MeV mass gap prohibits the $X(4500)$ and $X(4700)$ as the same doublet, and only the $X(4700)$ is favored as the $2S$ states in the present work.

From Fig.~\ref{tetra}, it can be seen that the $X(4140)$ is a good candidate of the lowest $1^{++}$ state. The predicted mass of the lowest $1^{++}$ state is 4195 MeV, which is the same value as that obtained in compact tetraquark picture by Stancu~\cite{Stancu:2009ka}. In the compact tetraquark scenario, there also exist a higher $1^{++}$ state, which may be assigned as the $X(4274)$~\cite{Stancu:2009ka}. By considering the $[6_c]_{cs}$ type diquarks, the diquark and antidiquark picture can also give the good description of the $X(4274)$~\cite{Chen:2016oma}. However, as emphasized in Sec.~\ref{diquark}, the $[6_c]_{cs}$ type diquarks can not be formed due to the repulsive confinement in the GI  quark model. Only one $1S$ $1^{++}$ state exists in the $[\bar 3_c]_{cs}\otimes [3_c]_{\bar c \bar s}$ diquark-antidiquark picture, and no room is left for the $X(4274)$~\cite{Lebed:2016yvr,Maiani:2016wlq}. Maiani et al., also discuss the possibility that the $X(4274)$ may have $0^{++}$ or $2^{++}$ quantum numbers~\cite{Maiani:2016wlq}. If so, the $X(4274)$ seems to be the candidate of the $1S$ $2^{++}$ state with 4302 MeV in the present work.
In addition, the obtained masses of the $1P$ tetraquark states are above 4.5 GeV. The four $1^{--}$ states are all around 4.6 GeV, which may correspond to the $X(4630)$ and $X(4660)$. The $1D$ tetraquarks are above 4.9 MeV, which are much higher than the energy region we are concerning.

To sum up, by considering the tetraquarks composed of $S$ and $A$ diquarks, we assign the $X(4140)$ as the $1S$ $1^{++}$ $cs\bar c \bar s$ tetraquark state and the $X(4700)$ as the $2S$ $0^{++}$ $cs\bar c \bar s$ tetraquark state within $S\bar S$ or $A\bar A$ type. There is no room left for the $X(4274)$ and $X(4500)$ in our calculation.

\subsection{Tetraquarks composed of internal excited diquarks}

Besides the $S$ and $A$ type diquarks, the internal excited diquarks have also been considered to account for the newly observed structures ~\cite{Chen:2016oma,Wang:2016gxp}. We adopt the diquark masses listed in Tab~\ref{tab1} to calculate the $0^{++}$ and $1^{++}$ tetraquark states. The results are presented in Tab.~\ref{mass2}.

\begin{table*}[!htbp]
\begin{center}
\caption{ \label{mass2} Masses of $cs\bar c \bar s$ tetraquark states composed of of internal excited diquarks. When the diquark and antidiquark are in different types, the linear combinations are understood to form the eigenstates of charge conjugation~\cite{Lebed:2016yvr}. The unit is in MeV.}
\small
\begin{tabular*}{17cm}{@{\extracolsep{\fill}}*{8}{p{1.7cm}<{\centering}}}
\hline\hline
 Diquark        & Antidiquark          & $n+1$           & $S$     & $L$     & Mass (GI model)  & Mass ($\mu=0.02~\rm{GeV}$) &  Mass ($\mu=0.04~\rm{GeV}$) \\\hline
                 $\bold {|0^{++}\rangle}$      &          &           &      &      &  & &   \\
 $2^1S_0$    & $\bar S$             & 1           & 0     & 0     & 4558       & 4516          & 4475    \\
 $2^3S_1$    & $\bar A$             & 1           & 0     & 0     & 4358       & 4315          & 4271    \\
 $1^3D_1$    & $\bar A$             & 1           & 0     & 0     & 4456       & 4410          & 4363    \\
 $1^1P_1$    & $\overline{1^1P_1}$             & 1           & 0     & 0     & 4490       & 4448          & 4405    \\
 $1^1P_1$    & $\overline{1^3P_1}$             & 1           & 0     & 0     & 4496       & 4453          & 4409    \\
 $1^3P_0$    & $\overline{1^3P_0}$             & 1           & 0     & 0     & 4727       & 4682          & 4639    \\
 $1^3P_1$    & $\overline{1^3P_1}$             & 1           & 0     & 0     & 4501       & 4458          & 4413    \\
 $1^1P_1$    & $\bar S$             & 1           & 1     & 1     & 4875       & 4837          & 4799    \\
 $1^1P_1$    & $\bar A$             & 1           & 1     & 1     & 4842       & 4805          & 4767    \\
 $1^3P_0$    & $\bar A$             & 1           & 1     & 1     & 4870       & 4831          & 4793    \\
 $1^3P_1$    & $\bar S$             & 1           & 1     & 1     & 4880       & 4842          & 4803    \\
 $1^3P_1$    & $\bar A$             & 1           & 1     & 1     & 4848       & 4810          & 4771    \\\hline
   $\bold {|1^{++}\rangle}$      &          &           &      &      &  & &   \\
 $2^1S_0$    & $\bar A$             & 1           & 1     & 0     & 4589       & 4547          & 4505    \\
 $2^3S_1$    & $\bar S$             & 1           & 1     & 0     & 4577       & 4534          & 4492    \\
 $2^3S_1$    & $\bar A$             & 1           & 1     & 0     & 4505       & 4461          & 4418    \\
 $1^3D_1$    & $\bar S$             & 1           & 1     & 0     & 4671       & 4626          & 4580    \\
 $1^3D_1$    & $\bar A$             & 1           & 1     & 0     & 4600       & 4554          & 4508    \\
 $1^1P_1$    & $\overline{1^3P_0}$             & 1           & 1     & 0     & 4732       & 4689          & 4646    \\
 $1^1P_1$    & $\overline{1^3P_1}$             & 1           & 1     & 0     & 4640       & 4598          & 4554    \\
 $1^3P_0$    & $\overline{1^3P_1}$             & 1           & 1     & 0     & 4738       & 4693          & 4649    \\
 $1^1P_1$    & $\bar S$             & 1           & 1     & 1     & 4892       & 4855          & 4816    \\
 $1^1P_1$    & $\bar A$             & 1           & 0     & 1     & 4923       & 4885          & 4847    \\
 $1^1P_1$    & $\bar A$             & 1           & 1     & 1     & 4924       & 4886          & 4847    \\
 $1^1P_1$    & $\bar A$             & 1           & 2     & 1     & 4845       & 4808          & 4770    \\
 $1^3P_0$    & $\bar S$             & 1           & 0     & 1     & 4902       & 4863          & 4824    \\
 $1^3P_0$    & $\bar A$             & 1           & 1     & 1     & 4904       & 4864          & 4825    \\
 $1^3P_1$    & $\bar S$             & 1           & 1     & 1     & 4898       & 4859          & 4820    \\
 $1^3P_1$    & $\bar A$             & 1           & 0     & 1     & 4929       & 4890          & 4850    \\
 $1^3P_1$    & $\bar A$             & 1           & 1     & 1     & 4929       & 4890          & 4850    \\
 $1^3P_2$    & $\bar A$             & 1           & 2     & 1     & 4850       & 4813          & 4774    \\
\hline\hline
\end{tabular*}
\end{center}
\end{table*}

For the $0^{++}$ tetraquark states, there are several combinations around 4.5 GeV, such as one $2S$ diquark and one $\bar S$ type antidiquark, and one $P$ wave excited diquark and one $P$ wave antidiquark. In Ref.~\cite{Chen:2016oma}, the $X(4500)$ is explained as one $D$ wave diquark and one $S$ wave antidiquark. In our present calculation, the central value of one $D$ wave diquark and one $S$ wave antidiquark case is 4410 MeV, which seems to be smaller than the experimental data. We, therefore,  prefer to interpret the $X(4500)$ as the tetraquark state composed of one $2^1S_0$ diquark and one $\bar S$ type antidiquark with the predicted mass 4516 MeV, although the one $P$ wave diquark and one $P$ wave antidiquark type tetraquark state cannot be simply excluded. It is also possible to explain the $X(4700)$ as the $^3P_0$ diquark-antidiquark state, while the ground tetraquark state with vector-diquark vector-antidiquark is favored in Ref.~\cite{Wang:2016gxp}.

The lowest state is the tetraquark composed of one $2^3S_1$ diquark and one $A$ type antidiquark with mass 4315 MeV. As emphasized in Sec. I, a narrow $J/\psi \phi$ peak $X(4350)$ in the double photon collisions has been observed~\cite{Shen:2009vs}, which indicates the $X(4350)$ should have the spin parity $J^{PC}=0^{++}$ or $2^{++}$. We believe that the $X(4350)$ is a good candidate of the lowest $0^{++}$ tetraquark with internal excited diquarks. In this scheme, the $\chi_{c0}(3915)$ is the ground $A\bar A$ tetraquark state, the $X(4700)$ is the $2S$ $A\bar A$ or $S\bar S$ state, the $X(4350)$ is the $2^3S_1\bar A$ tetraquark state, and the $X(4500)$ is the $2^1S_0 \bar S$ tetraquark state. The absent $0^{++}$ one without orbital excitation is the ground $S\bar S$ tetraquark state. The $X(4140)$ is also the ground state $1^{++}$ $A \bar S$ tetraquark state without orbital excitation. The predicted $1^{++}$ tetraquark states with internal excited diquarks are much higher than the $X(4274)$, which disfavors the $X(4274)$ as the $1^{++}$ tetraquark state.

\subsection{$P$ wave charmonium}

\begin{figure}[!htbp]
\includegraphics[scale=0.52]{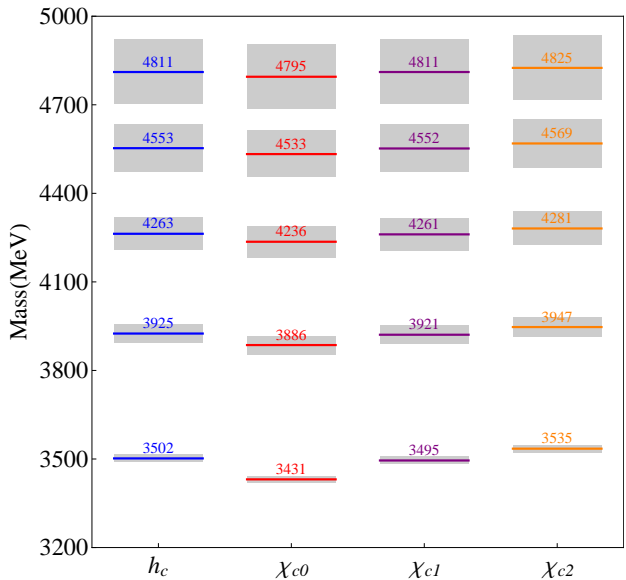}
\vspace{0.0cm} \caption{Mass of the $P$ wave charmonium up to 5 GeV.}
\label{charmonium}
\end{figure}

According to the Particle Data Group~\cite{Agashe:2014kda}, the four $1P$ charmonium states and $\chi_{c2}(2P)$ have been well established. The assignment of the $\chi_{c0}(3915)$ is in debet. The $\chi_{c1}(2^3P_1)$ state is related with the nature of the mysterious $X(3872)$, which can hardly be solved in the near future. The mass spectrum and strong decay behavior of the $3P$ states are also investigated in the Refs.~\cite{Barnes:2005pb,Liu:2009fe}. Here, we present the mass spectrum of $P$ charmonium up to 5 GeV in Fig.~\ref{charmonium}.

The predicted mass of the $\chi_{c1}(3^3P_1)$ state is 4261 MeV, in good agreement with the $X(4274)$. With the quark pair creation model, the total decay width of the $\chi_{c1}(3^3P_1)$ is 39 MeV~\cite{Barnes:2005pb}, which is consistent with the experimental width $56\pm11^{+8}_{-11}~\rm{MeV}$. In the relativized quark model, it can be seen that for the $1^{++}$ states around 4.2 GeV, one tetraquark state corresponds to the $X(4140)$, and the charmonium stands for the $X(4274)$. More arguments can be found in Ref.~\cite{Liu:2016onn}, where the $X(4274)$ is suggested as the $\chi_{c1}(3^3P_1)$ state.

We notice that the $X(4500)$ and $X(4700)$ also lie in the energy regions of the $\chi_{c0}(4^3P_0)$ and $\chi_{c0}(5^3P_0)$. The screening  effects are small for the low lying charmonium, while the uncertainties become much larger for the $4P$ and $5P$ states. These uncertainties of mass spectra prohibit us to give a reliable conclusion on the $\chi_{c0}(4^3P_0)$ and $\chi_{c0}(5^3P_0)$ states. Other information, such as the strong and radiative decay behaviors of these higher charmonium, are needed. Moreover, under the factorization ansatz, the $1^{++}$ charmonium can be produced more easily than the $0^{++}$ charmonium in the $B^+ \to K^+ + X$ process~\cite{Liu:2016onn}, which indicates the $1^{++}$ $X(4274)$ is suitable for the conventional charmonium, but the $0^{++}$ $X(4500)$ and $X(4700)$ not.

\section{Summary}{\label{Summary}}

In this work, we investigate the masses of $cs \bar c \bar s$ tetraquark states in the diquark-antidiquark picture using the relativized quark model proposed by Godfrey and Isgur. The diquark and antidiquark masses are obtained with the relativized potential firstly, and then the diquark and antidiquark are treated as the usual
point-like antiquark and quark, respectively. Here, only the antitriplet diquark $[\bar 3_c]_{cs}$ is considered. The masses of tetraquark states are obtained by solving the Schr\"{o}dinger-type equation between diquark and antidiquark. The color screening effects are also added in the present work.

It is found that the resonance $X(4140)$ can be regarded as the $A\bar S$ type tetraquark state, and the $X(4700)$ can be assigned as the the $2S$ $A\bar A$ or $S\bar S$ state. When the internal excited diquarks are considered, the $X(4500)$ can be explained as the tetraquark composed of one $2S$ scalar diquark and one scalar antidiquark. The $X(4274)$ as the tetraquark state is not favored, however, it could be a good candidate of the conventional $\chi_{c1}(3^3P_1)$ state. Other charmonium-like states $\chi_{c0}(3915)$, $X(4350)$, $X(4630)$ and $X(4660)$ are also discussed in this work. We expect our assignments can be tested by future experiments.

{\bf\it Acknowledgement:}  We would like to thank Zong-Ye Zhang, De-Min Li, and Elena Santopinto for valuable discussions. This project is supported by the National Natural Science Foundation of China under Grants No.~10975146, and No.~11475192. The fund provided by the Sino-German CRC 110 ``Symmetries and the Emergence of Structure in QCD" project (NSFC Grant No. 11261130311) is also appreciated.


\begin{thebibliography}{120}

\bibitem{Aaij:2016iza}
  R.~Aaij {\it et al.} (LHCb Collaboration),
  Observation of $J/\psi\phi$ structures consistent with exotic states from amplitude analysis of $B^+\to J/\psi \phi K^+$ decays,
  arXiv:1606.07895.

\bibitem{Aaij:2016nsc}
  R.~Aaij {\it et al.} (LHCb Collaboration),
  Amplitude analysis of $B^+\to J/\psi \phi K^+$ decays,
  arXiv:1606.07898.

\bibitem{Aaltonen:2009tz}
  T.~Aaltonen {\it et al.} (CDF Collaboration),
  Evidence for a Narrow Near-Threshold Structure in the $J/\psi\phi$ Mass Spectrum in $B^+\to J/\psi\phi K^+$ Decays,
  Phys.\ Rev.\ Lett.\  {\bf 102}, 242002 (2009).

\bibitem{Brodzicka:2010zz}
  J.~Brodzicka,
  Heavy Flavour Spectroscopy,
  Conf.\ Proc.\ C {\bf 0908171}, 299 (2009).

\bibitem{Aaltonen:2011at}
  T.~Aaltonen {\it et al.} (CDF Collaboration),
  Observation of the $Y(4140)$ structure in the $J/\psi\,\phi$ Mass Spectrum in $B^\pm\to J/\psi\,\phi K$ decays,
  arXiv:1101.6058.

\bibitem{Aaij:2012pz}
  R.~Aaij {\it et al.} (LHCb Collaboration),
  Search for the $X(4140)$ state in $B^+ \to J/\psi \phi K^+$ decays,
  Phys.\ Rev.\ D {\bf 85}, 091103 (2012).

\bibitem{Chatrchyan:2013dma}
  S.~Chatrchyan {\it et al.} (CMS Collaboration),
  Observation of a peaking structure in the $J/\psi \phi$ mass spectrum from $B^{\pm} \to J/\psi \phi K^{\pm}$ decays,
  Phys.\ Lett.\ B {\bf 734}, 261 (2014).

\bibitem{Abazov:2013xda}
  V.~M.~Abazov {\it et al.} (D0 Collaboration),
  Search for the $X(4140)$ state in $B^+ \to J/\psi \phi K^+$ decays with the D0 detector,
  Phys.\ Rev.\ D {\bf 89}, 012004 (2014).

\bibitem{Lees:2014lra}
  J.~P.~Lees {\it et al.} (BaBar Collaboration),
  Study of $B^{\pm,0} \to J/\psi K^+ K^- K^{\pm,0}$ and search for $B^0 \to J/\psi\phi$ at BABAR,
  Phys.\ Rev.\ D {\bf 91}, 012003 (2015).

\bibitem{Abazov:2015sxa}
  V.~M.~Abazov {\it et al.} (D0 Collaboration),
  Inclusive Production of the $X(4140)$ State in $p \overline p$ Collisions at D0,
  Phys.\ Rev.\ Lett.\  {\bf 115}, 232001 (2015).

\bibitem{Shen:2009vs}
  C.~P.~Shen {\it et al.} (Belle Collaboration),
  Evidence for a new resonance and search for the $Y(4140)$ in the $\gamma \gamma \to \phi J/\psi$ process,
  Phys.\ Rev.\ Lett.\  {\bf 104}, 112004 (2010).

\bibitem{Liu:2009ei}
  X.~Liu and S.~L.~Zhu,
  $Y(4143)$ is probably a molecular partner of $Y(3930)$,
  Phys.\ Rev.\ D {\bf 80}, 017502 (2009); Erratum: [Phys.\ Rev.\ D {\bf 85}, 019902 (2012)].

\bibitem{Mahajan:2009pj}
  N.~Mahajan,
  $Y(4140)$: Possible options,
  Phys.\ Lett.\ B {\bf 679}, 228 (2009).

\bibitem{Wang:2009ue}
  Z.~G.~Wang,
  Analysis of the $Y(4140)$ with QCD sum rules,
  Eur.\ Phys.\ J.\ C {\bf 63}, 115 (2009).

\bibitem{Wang:2009ry}
  Z.~G.~Wang, Z.~C.~Liu and X.~H.~Zhang,
  Analysis of the $Y(4140)$ and related molecular states with QCD sum rules,
  Eur.\ Phys.\ J.\ C {\bf 64}, 373 (2009).



\bibitem{Albuquerque:2009ak}
  R.~M.~Albuquerque, M.~E.~Bracco and M.~Nielsen,
  A QCD sum rule calculation for the $Y(4140)$ narrow structure,
  Phys.\ Lett.\ B {\bf 678}, 186 (2009).

\bibitem{Branz:2009yt}
  T.~Branz, T.~Gutsche and V.~E.~Lyubovitskij,
  Hadronic molecule structure of the $Y(3940)$ and $Y(4140)$,
  Phys.\ Rev.\ D {\bf 80}, 054019 (2009).

\bibitem{Ding:2009vd}
  G.~J.~Ding,
  Possible Molecular States of $D^*_s$ $\bar D^*_s$ System and $Y(4140)$,
  Eur.\ Phys.\ J.\ C {\bf 64}, 297 (2009).

\bibitem{Zhang:2009st}
  J.~R.~Zhang and M.~Q.~Huang,
  $(Q\bar s)^{(*)}(\bar Q s)^{(*)}$ molecular states from QCD sum rules: A view on $Y(4140)$,
  J.\ Phys.\ G {\bf 37}, 025005 (2010).


\bibitem{Shen:2010ky}
  L.~L.~Shen, X.~L.~Chen, Z.~G.~Luo, P.~Z.~Huang, S.~L.~Zhu, P.~F.~Yu and X.~Liu,
  The Molecular systems composed of the charmed mesons in the $H\bar{S}+h.c.$ doublet,
  Eur.\ Phys.\ J.\ C {\bf 70}, 183 (2010).

\bibitem{Liu:2010hf}
  X.~Liu, Z.~G.~Luo and S.~L.~Zhu,
  Novel charmonium-like structures in the $J/\psi\phi$ and $J/\psi\omega$ invariant mass spectra,
  Phys.\ Lett.\ B {\bf 699}, 341 (2011);Erratum: [Phys.\ Lett.\ B {\bf 707}, 577 (2012)].

\bibitem{Finazzo:2011he}
  S.~I.~Finazzo, M.~Nielsen and X.~Liu,
  QCD sum rule calculation for the charmonium-like structures in the $J/\psi \phi$ and $J/\psi \omega$ invariant mass spectra,
  Phys.\ Lett.\ B {\bf 701}, 101 (2011).


\bibitem{He:2011ed}
  J.~He and X.~Liu,
  The open-charm radiative and pionic decays of molecular charmonium $Y(4274)$,
  Eur.\ Phys.\ J.\ C {\bf 72}, 1986 (2012).

\bibitem{HidalgoDuque:2012pq}
  C.~Hidalgo-Duque, J.~Nieves and M.~P.~Valderrama,
  Light flavor and heavy quark spin symmetry in heavy meson molecules,
  Phys.\ Rev.\ D {\bf 87}, 076006 (2013).


\bibitem{Wang:2014gwa}
  Z.~G.~Wang,
  Reanalysis of the $Y(3940)$, $Y(4140)$, $Z_c(4020)$, $Z_c(4025)$ and $Z_b(10650)$ as molecular states with QCD sum rules,
  Eur.\ Phys.\ J.\ C {\bf 74}, no. 7, 2963 (2014).

\bibitem{Ma:2014ofa}
  L.~Ma, Z.~F.~Sun, X.~H.~Liu, W.~Z.~Deng, X.~Liu and S.~L.~Zhu,
  Probing the $XYZ$ states through radiative decays,
  Phys.\ Rev.\ D {\bf 90}, 034020 (2014).

\bibitem{Ma:2014zva}
  L.~Ma, X.~H.~Liu, X.~Liu and S.~L.~Zhu,
  Strong decays of the $XYZ$ states,
  Phys.\ Rev.\ D {\bf 91}, 034032 (2015).

\bibitem{Karliner:2016ith}
  M.~Karliner and J.~L.~Rosner,
  Exotic resonances due to $\eta$ exchange,
  arXiv:1601.00565.

\bibitem{Torres:2016oyz}
  A.~Mart¨ªnez Torres, K.~P.~Khemchandani, J.~M.~Dias, F.~S.~Navarra and M.~Nielsen,
  Understanding close-lying exotic charmonia states within QCD sum rules,
  arXiv:1606.07505.




\bibitem{Stancu:2009ka}
  F.~Stancu,
  Can Y(4140) be a $c \bar c s \bar s$ tetraquark?,
  J.\ Phys.\ G {\bf 37}, 075017 (2010).

\bibitem{Drenska:2009cd}
  N.~V.~Drenska, R.~Faccini and A.~D.~Polosa,
  Exotic Hadrons with Hidden Charm and Strangeness,
  Phys.\ Rev.\ D {\bf 79}, 077502 (2009).

\bibitem{Ebert:2010zz}
  D.~Ebert, R.~N.~Faustov and V.~O.~Galkin,
  Masses of heavy tetraquarks with hidden charm and bottom,
  Phys.\ Part.\ Nucl.\  {\bf 41}, 931 (2010).


\bibitem{Chen:2010ze}
  W.~Chen and S.~L.~Zhu,
  The Vector and Axial-Vector Charmonium-like States,
  Phys.\ Rev.\ D {\bf 83}, 034010 (2011).

\bibitem{Vijande:2014cfa}
  J.~Vijande and A.~Valcarce,
  Unraveling the pattern of the XYZ mesons,
  Phys.\ Lett.\ B {\bf 736}, 325 (2014).

\bibitem{Patel:2014vua}
  S.~Patel, M.~Shah and P.~C.~Vinodkumar,
  Mass spectra of four-quark states in the hidden charm sector,
  Eur.\ Phys.\ J.\ A {\bf 50}, 131 (2014).

\bibitem{Anisovich:2015caa}
  V.~V.~Anisovich, M.~A.~Matveev, A.~V.~Sarantsev and A.~N.~Semenova,
  Exotic mesons with hidden charm as diquark¨Cantidiquark states,
  Int.\ J.\ Mod.\ Phys.\ A {\bf 30}, 1550186 (2015).


\bibitem{Wang:2015pea}
  Z.~G.~Wang and Y.~F.~Tian,
  Tetraquark state candidates: $Y$(4140), $Y$(4274) and $X$(4350),
  Int.\ J.\ Mod.\ Phys.\ A {\bf 30}, 1550004 (2015).

\bibitem{Zhou:2015frp}
  P.~Zhou, C.~R.~Deng and J.~L.~Ping,
  Identification of $Y (4008)$, $Y (4140)$, $Y (4260)$, and $Y(4360)$ as Tetraquark States,
  Chin.\ Phys.\ Lett.\  {\bf 32}, 101201 (2015).

\bibitem{Padmanath:2015era}
  M.~Padmanath, C.~B.~Lang and S.~Prelovsek,
  $X(3872)$ and $Y(4140)$ using diquark-antidiquark operators with lattice QCD,
  Phys.\ Rev.\ D {\bf 92}, 034501 (2015).

\bibitem{Lebed:2016yvr}
  R.~F.~Lebed and A.~D.~Polosa,
  $\chi_{c0}(3915)$ As the Lightest $c\bar c s \bar s$ State,
  Phys.\ Rev.\ D {\bf 93}, 094024 (2016).

\bibitem{vanBeveren:2009dc}
  E.~van Beveren and G.~Rupp,
  The $Y(4140)$, $X(4260)$, $\psi(2D)$, $\psi(4S)$ and tentative $\psi(3D)$,
  arXiv:0906.2278.

\bibitem{Swanson:2014tra}
  E.~S.~Swanson,
  $Z_b$ and $Z_c$ Exotic States as Coupled Channel Cusps,
  Phys.\ Rev.\ D {\bf 91}, 034009 (2015).

\bibitem{Molina:2009ct}
  R.~Molina and E.~Oset,
  The $Y(3940)$, $Z(3930)$ and the $X(4160)$ as dynamically generated resonances from the vector-vector interaction,
  Phys.\ Rev.\ D {\bf 80}, 114013 (2009).

\bibitem{Branz:2010rj}
  T.~Branz, R.~Molina and E.~Oset,
  Radiative decays of the $Y(3940)$, $Z(3930)$ and the $X(4160)$ as dynamically generated resonances,
  Phys.\ Rev.\ D {\bf 83}, 114015 (2011).

\bibitem{Liu:2009iw}
  X.~Liu,
  The Hidden charm decay of $Y(4140)$ by the rescattering mechanism,
  Phys.\ Lett.\ B {\bf 680}, 137 (2009).

\bibitem{Chen:2016qju}
  H.~X.~Chen, W.~Chen, X.~Liu and S.~L.~Zhu,
  The hidden-charm pentaquark and tetraquark states,
  Phys.\ Rept.\  {\bf 639}, 1 (2016).

\bibitem{Chen:2016oma}
  H.~X.~Chen, E.~L.~Cui, W.~Chen, X.~Liu and S.~L.~Zhu,
  Understanding the internal structures of the $X(4140)$, $X(4274)$, $X(4500)$ and $X(4700)$,
  arXiv:1606.03179.

\bibitem{Wang:2016gxp}
  Z.~G.~Wang,
  Scalar tetraquark state candidates: $X(3915)$, $X(4500)$ and $X(4700)$,
  arXiv:1606.05872.

\bibitem{Wang:2016tzr}
  Z.~G.~Wang,
  Reanalysis of the $X(4140)$ as axialvector tetraquark state with QCD sum rules,
  arXiv:1607.00701.

\bibitem{Liu:2016onn}
  X.~H.~Liu,
  How to understand the underlying structures of $X(4140)$, $X(4274)$, $X(4500)$ and $X(4700)$,
  arXiv:1607.01385.

\bibitem{Maiani:2016wlq}
  L.~Maiani, A.~D.~Polosa and V.~Riquer,
  Interpretation of Axial Resonances in $J/\psi \phi$ at LHCb,
  arXiv:1607.02405.

\bibitem{Zhu:2016arf}
  R.~Zhu,
  Hidden charm octet tetraquarks from a diquark-antidiquark model,
  arXiv:1607.02799.

\bibitem{Ali:2016dkf}
  A.~Ali, I.~Ahmed, M.~J.~Aslam and A.~Rehman,
  Heavy quark symmetry and weak decays of the $b$-baryons in pentaquarks with a $c\bar{c}$ component,
  arXiv:1607.00987.

\bibitem{He:2016pfa}
  J.~He,
  $P$-wave dynamical generated state and LHCb hidden-charmed pentaquarks,
  arXiv:1607.03223.

\bibitem{Wang:2016ujn}
  Z.~G.~Wang,
  Reanalysis of the $X(3915)$, $X(4500)$ and $X(4700)$ with QCD sum rules,
  arXiv:1607.04840.

\bibitem{Godfrey:1985xj}
  S.~Godfrey and N.~Isgur,
  Mesons in a Relativized Quark Model with Chromodynamics,
  Phys.\ Rev.\ D {\bf 32}, 189 (1985).

\bibitem{Capstick:1986bm}
  S.~Capstick and N.~Isgur,
  Baryons in a Relativized Quark Model with Chromodynamics,
  Phys.\ Rev.\ D {\bf 34}, 2809 (1986).

\bibitem{Godfrey:1998pd}
  S.~Godfrey and J.~Napolitano,
  Light meson spectroscopy,
  Rev.\ Mod.\ Phys.\  {\bf 71}, 1411 (1999).

\bibitem{Ferretti:2013faa}
  J.~Ferretti, G.~Galat\`a and E.~Santopinto,
  Interpretation of the $X(3872)$ as a charmonium state plus an extra component due to the coupling to the meson-meson continuum,
  Phys.\ Rev.\ C {\bf 88}, 015207 (2013).

\bibitem{Ferretti:2013vua}
  J.~Ferretti and E.~Santopinto,
  Higher mass bottomonia,
  Phys.\ Rev.\ D {\bf 90}, 094022 (2014).


\bibitem{Godfrey:2015dva}
  S.~Godfrey and K.~Moats,
  Properties of Excited Charm and Charm-Strange Mesons,
  Phys.\ Rev.\ D {\bf 93}, 034035 (2016).



\bibitem{Capstick:1992th}
  S.~Capstick and W.~Roberts,
  $N \pi$ decays of baryons in a relativized model,
  Phys.\ Rev.\ D {\bf 47}, 1994 (1993).


\bibitem{Barnes:2005pb}
  T.~Barnes, S.~Godfrey and E.~S.~Swanson,
  Higher charmonia,
  Phys.\ Rev.\ D {\bf 72}, 054026 (2005).

\bibitem{Liu:2009fe}
  X.~Liu, Z.~G.~Luo and Z.~F.~Sun,
  $X(3915)$ and $X(4350)$ as new members in $P$-wave charmonium family,
  Phys.\ Rev.\ Lett.\  {\bf 104}, 122001 (2010).

\bibitem{Zhou:2013ada}
  Z.~Y.~Zhou and Z.~Xiao,
  Comprehending heavy charmonia and their decays by hadron loop effects,
  Eur.\ Phys.\ J.\ A {\bf 50}, 165 (2014).

\bibitem{Lu:2014zua}
  Q.~F.~L\"{u} and D.~M.~Li,
  Understanding the charmed states recently observed by the LHCb and BaBar Collaborations in the quark model,
  Phys.\ Rev.\ D {\bf 90}, 054024 (2014).

\bibitem{Song:2015nia}
  Q.~T.~Song, D.~Y.~Chen, X.~Liu and T.~Matsuki,
  Charmed-strange mesons revisited: mass spectra and strong decays,
  Phys.\ Rev.\ D {\bf 91}, 054031 (2015).

\bibitem{Song:2015fha}
  Q.~T.~Song, D.~Y.~Chen, X.~Liu and T.~Matsuki,
  Higher radial and orbital excitations in the charmed meson family,
  Phys.\ Rev.\ D {\bf 92}, 074011 (2015).

\bibitem{Godfrey:2014fga}
  S.~Godfrey and K.~Moats,
  The $D_{sJ}^*(2860)$ Mesons as Excited D-wave $c\bar{s}$ States,
  Phys.\ Rev.\ D {\bf 90}, 117501 (2014).


\bibitem{Godfrey:2015dia}
  S.~Godfrey and K.~Moats,
  Bottomonium Mesons and Strategies for their Observation,
  Phys.\ Rev.\ D {\bf 92}, 054034 (2015).

\bibitem{Ferretti:2015rsa}
  J.~Ferretti and E.~Santopinto,
  Open-flavor strong decays of open-charm and open-bottom mesons in the $^3P_0$ pair-creation model,
  arXiv:1506.04415.

\bibitem{Capstick:1989ra}
  S.~Capstick and S.~Godfrey,
  Pseudoscalar Decay Constants in the Relativized Quark Model and Measuring the {CKM} Matrix Elements,
  Phys.\ Rev.\ D {\bf 41}, 2856 (1990).

\bibitem{Yaouanc:2014jka}
  H.-R.~Dong, A.~Le Yaouanc, L.~Oliver, and J.-C.~Raynal,
  Finite mass corrections for $B \to( \bar D^{(*)},\bar D^{**})\ell \nu$ decays in the Bakamjian-Thomas relativistic quark model,
  Phys.\ Rev.\ D {\bf 90}, 114014 (2014).

\bibitem{Godfrey:2016nwn}
  S.~Godfrey, K.~Moats and E.~S.~Swanson,
  $B$ and $B_s$ Meson Spectroscopy,
  arXiv:1607.02169.

\bibitem{Ebert:2002ig}
  D.~Ebert, R.~N.~Faustov, V.~O.~Galkin and A.~P.~Martynenko,
  Mass spectra of doubly heavy baryons in the relativistic quark model,
  Phys.\ Rev.\ D {\bf 66}, 014008 (2002).

\bibitem{Ebert:2005nc}
  D.~Ebert, R.~N.~Faustov and V.~O.~Galkin,
  Masses of heavy tetraquarks in the relativistic quark model,
  Phys.\ Lett.\ B {\bf 634}, 214 (2006).



\bibitem{Ebert:2008kb}
  D.~Ebert, R.~N.~Faustov and V.~O.~Galkin,
  Excited heavy tetraquarks with hidden charm,
  Eur.\ Phys.\ J.\ C {\bf 58}, 399 (2008).

\bibitem{Ebert:2010af}
  D.~Ebert, R.~N.~Faustov and V.~O.~Galkin,
  Masses of tetraquarks with open charm and bottom,
  Phys.\ Lett.\ B {\bf 696}, 241 (2011)

\bibitem{Santopinto:2014opa}
  E.~Santopinto and J.~Ferretti,
  Strange and nonstrange baryon spectra in the relativistic interacting quark-diquark model with a Grsey and Radicati-inspired exchange interaction,
  Phys.\ Rev.\ C {\bf 92}, 025202 (2015).


\bibitem{Hadizadeh:2015cvx}
  M.~R.~Hadizadeh and A.~Khaledi-Nasab,
  Heavy tetraquarks in the diquark-antidiquark picture,
  Phys.\ Lett.\ B {\bf 753}, 8 (2016).


\bibitem{D0:2016mwd}
  V.~M.~Abazov {\it et al.} (D0 Collaboration),
  Evidence for a $B_s^0 \pi^\pm$ state,
  Phys.\ Rev.\ Lett.\ {\bf 117}, 022003 (2016).

\bibitem{LHCb:2016ppf}
  The LHCb Collaboration (LHCb Collaboration),
  Search for structure in the $B_s^0\pi^\pm$ invariant mass spectrum,
  LHCb-CONF-2016-004, CERN-LHCb-CONF-2016-004.


\bibitem{Lu:2016zhe}
  Q.~F.~L\"u and Y.~B.~Dong,
  Masses of open charm and bottom tetraquark states in relativized quark model,
  arXiv:1603.06417.

\bibitem{Dong:1994zj}
  Y.~B.~Dong, Y.~W.~Yu, Z.~Y.~Zhang and P.~N.~Shen,
  Leptonic decay of charmonium,
  Phys.\ Rev.\ D {\bf 49}, 1642 (1994).

\bibitem{Ding:1995he}
  Y.~B.~Ding, K.~T.~Chao and D.~H.~Qin,
  Possible effects of color screening and large string tension in heavy quarkonium spectra,
  Phys.\ Rev.\ D {\bf 51}, 5064 (1995).

\bibitem{Hiyama:2003cu}
  E.~Hiyama, Y.~Kino and M.~Kamimura,
  Gaussian expansion method for few-body systems,
  Prog.\ Part.\ Nucl.\ Phys.\  {\bf 51}, 223 (2003).


\bibitem{Maiani:2004vq}
  L.~Maiani, F.~Piccinini, A.~D.~Polosa and V.~Riquer,
  Diquark-antidiquarks with hidden or open charm and the nature of X(3872),
  Phys.\ Rev.\ D {\bf 71}, 014028 (2005).


\bibitem{Santopinto:2004hw}
  E.~Santopinto,
  An Interacting quark-diquark model of baryons,
  Phys.\ Rev.\ C {\bf 72}, 022201 (2005).

\bibitem{Ferretti:2011zz}
  J.~Ferretti, A.~Vassallo and E.~Santopinto,
  Relativistic quark-diquark model of baryons,
  Phys.\ Rev.\ C {\bf 83}, 065204 (2011).


\bibitem{Maiani:2014aja}
  L.~Maiani, F.~Piccinini, A.~D.~Polosa and V.~Riquer,
  The Z(4430) and a New Paradigm for Spin Interactions in Tetraquarks,
  Phys.\ Rev.\ D {\bf 89}, 114010 (2014).

\bibitem{Maiani:2015vwa}
  L.~Maiani, A.~D.~Polosa and V.~Riquer,
  The New Pentaquarks in the Diquark Model,
  Phys.\ Lett.\ B {\bf 749}, 289 (2015).

\bibitem{Brodsky:2014xia}
  S.~J.~Brodsky, D.~S.~Hwang and R.~F.~Lebed,
  Dynamical Picture for the Formation and Decay of the Exotic XYZ Mesons,
  Phys.\ Rev.\ Lett.\  {\bf 113}, 112001 (2014).

\bibitem{Lebed:2015tna}
  R.~F.~Lebed,
  The Pentaquark Candidates in the Dynamical Diquark Picture,
  Phys.\ Lett.\ B {\bf 749}, 454 (2015).

\bibitem{Agashe:2014kda}
  K.~A.~Olive {\it et al.} (Particle Data Group Collaboration),
  Review of Particle Physics,
  Chin.\ Phys.\ C {\bf 38}, 090001 (2014).




\end{thebibliography}
\end{document}